\documentclass{article}
\usepackage[english]{babel}
\usepackage[T2A]{fontenc}

\usepackage[tbtags]{amsmath}
\usepackage{amsfonts,amssymb,mathrsfs,amscd}
\usepackage{tikz}
\usepackage{hyperref}
\usepackage{lscape}

\numberwithin{equation}{section}
\begin{document}
    \title{Dynamical space-time ray tracing and\\  modified horizontal ray method}
    \author{A. Kaplun\footnote{University of Haifa,\,\,\href{mailto: alex.v.kaplun@gmail.com}{alex.v.kaplun@gmail.com},\,\,\href{mailto: a.kaplun.work@gmail.com}{a.kaplun.work@gmail.com}}, B. Katsnelson\footnote{University of Haifa,\,\,\href{mailto: bkatsnels@univ.haifa.ac.il}{bkatsnels@univ.haifa.ac.il}}}
\date{}

    \maketitle

\begin{abstract}
The "vertical modes and horizontal rays" method, commonly applied for simulating acoustic wave propagation in shallow water, as presented in \cite{burridge2005horizontal, connor1974complex, weinberg1974horizontal}, is advanced in this research. Our approach to this method involves the use of the so-called space-time rays \cite{babich1998space, babich1980space}, which are constructed by decomposing the time-dependent sound field into adiabatic vertical modes, the solutions to the Sturm-Liouville problem.  The introduction of the time coordinate, while still considering it as an additional space coordinate instead of merely a parameter along the ray, allows us to describe the propagation of frequency-modulated signals in an effectively frequency-dispersive medium. The consideration of the extension of Hamiltonian ray-tracing methods \cite{bergman2005generalized, vcerveny1982space, farra1993ray, farra1989ray, kirpichnikova1983reflection} (also used for the description of Gaussian beams and so-called quasiphotons \cite{babich1984complex, belishev1996boundary, kiselev2012nonparaxial, perel2021quasiphotons}) leads to a simple description of observable effects such as changes in modulation, time compression, differences between angles of phase and amplitude fronts, space-time caustics, etc., in dynamics — on the moving line or at some point of observation while having the general form of the source (for example, also a moving one).

\end{abstract}
\section{Introduction}
This paper aims to fill the gap existing between two approaches to the solution of the problem of acoustic wave propagation in a shallow sea. On the one hand, there is a representation for the solution of the wave equation 
\begin{align}
    & \Delta u - \frac{1}{c^2}\frac{\partial^2  }{\partial t^2} u  = 0
\end{align}
in the form of decomposition by vertical modes - eigenfunctions along the vertical axis \cite{burridge2005horizontal,felsen1981hybrid, weinberg1974horizontal}:
\begin{align}
    & u(t,\mathbf{r},z)\sim \int \sum_{n} a_n(\mathbf{r},\omega) \psi_n (z,\mathbf{r},\omega)\exp(- i \omega t) d \omega.
\end{align}
This approach provides an adequate description of the field under conditions of frequency variation in the small range $[\omega_0 - \delta\omega,\omega_0 + \delta\omega]$, but obtaining results in the case of substantial frequency modulation requires multiple solutions of the equations for different modes and frequencies, followed by the application of the Fourier transform \cite{katsnelson2018variability,katsnel2012space}. Therefore, it is difficult and resource-consuming to apply this method for the numerical calculation of wave process parameters in the general case. On the other hand, there is a method of considering of wave processes in the presence of frequency dispersion and modulation - space-time rays \cite{babich1980space,babich1984complex,bergman2005generalized,vcerveny1982space,vcerveny2020seismic,connor1974complex,kirpichnikova1983reflection} used in various fields of mathematics and physics. Such an object has previously been defined \cite{weinberg1974horizontal} for the study of horizontal refraction, but further development of the mathematical apparatus is required to study and predict the various observed effects (other than refraction proper).The particular feature of the proposed approach is also the fact that all observed effects are determined by the structure of the function $q_l$ (eigenvalue corresponding to the vertical mode), which contains all necessary information about possible dispersion effects; therefore, it does not require the construction of a separate model for each specific case (parameters of sound speed, bottom, obstacles, etc.). Detailed reviews of the two methods described above are given in \cite{babich1998space,etter2018underwater,katsnelson2012fundamentals}. Another basis of our research originates from the works considering the Hamiltonian approach for ray tracing \cite{farra1993ray,farra1989ray}, which allows us to study beam parameters (paraxial rays) from the general perspective.

It will also be important to note that our approach is close to the one used to solve the wave equation in a material medium with arbitrary dispersion \cite{kravtsov1990geometrical}:
\begin{align}
& \Delta u - \frac{1}{c^2}\frac{\partial^2 }{\partial t^2} \check{M}[u] = 0,
\end{align}
where $\check{M}$ is some integral operator describing the properties of the medium.

An additional motivation for this work is the fact that the developed mathematical apparatus turns out to be convenient for describing wave processes in other dispersive media as well. Several effects due to the influence of the dispersion in an inhomogeneous medium have been found in thepropagation of electromagnetic waves: in the optics of short, picosecond and femtosecond pulses \cite{bor1985group,hebling1996derivation,hebling2008generation,osvay2004angular,topp1975group}. In particular, these effects manifest themselves in the form of a tilt between the phase front and the pulse front after propagation through a dispersive medium (prism) \cite{bor1985group,hebling1996derivation,osvay2004angular,topp1975group} and are used in the nonlinear generation of terahertz pulses \cite{hebling2008generation}.
\section{Vertical modes and field distribution}
\subsection{Wave equation}
Let us consider the wave equation describing the propagation of acoustic waves in the sea:
\begin{align}
    & \left[\frac{\partial ^2}{\partial X^2} + \frac{\partial ^2}{\partial Y^2} + \frac{\partial ^2}{\partial Z^2} - \frac{1}{c(X,Y,Z)^2} \frac{\partial ^2}{\partial T^2}\right] u(T,X,Y,Z) = 0.
\end{align}
Here, $X,Y$ are horizontal coordinates and $Z\ge 0$ is vertical, with $Z=0$ being the equation of the water surface, $Z= h(X,Y)$ being the equation of the lower surface (bottom) at the point $(X,Y)$, and $c(X,Y,Z)$ being the speed of sound in the medium, which depends on the depth $Z$ significantly more than on the horizontal coordinates $(X,Y)$ \cite{weinberg1974horizontal}. The desired function $u$ has the physical meaning of acoustic pressure. Let us rewrite the equation in the form
\begin{align}
    & \left[- n^2(X,Y,Z) \frac{\partial ^2}{\partial (c_0 T)^2} +\frac{\partial ^2}{\partial X^2} + \frac{\partial ^2}{\partial Y^2} + \frac{\partial ^2}{\partial Z^2} \right] u(c_0 T,X,Y,Z) = 0,
\end{align}
where $c_0$ is some typical speed of wave propagation, and the following condition is satisfied:
\begin{align}
    & n(X,Y,Z) := \frac{c_0}{c(X,Y,Z)}\sim 1.
\end{align}
The boundary conditions for the wave equation are written in the form:
\begin{align}
    & u(T,X,Y,Z)|_{Z=0} = 0,\\
    & u (T,X,Y,Z)|_{Z=h(X,Y)+} = u (T,X,Y,Z)|_{Z=h(X,Y)-}\\
    & \frac{1}{\rho^{+}} \frac{\partial u (T,X,Y,Z)}{\partial Z}|_{Z=h(X,Y)+}  = \frac{1}{\rho^{-}} \frac{\partial u (T,X,Y,Z)}{\partial Z}|_{Z=h(X,Y)-}.
\end{align}
where $\rho^{+},\rho^{-}$ are the densities of the matter above and below the bottom, respectively, and we also require that the radiation conditions for the function $u$ are satisfied. We will assume that the following estimates relating to the derivatives for different variables are satisfied:
\begin{align}
    & \frac{\partial}{\partial (\varepsilon X)}  \sim \frac{\partial}{\partial (\varepsilon Y)}  \sim \frac{\partial}{\partial (\varepsilon c_0 T)} \sim \frac{\partial}{\partial Z}.
\end{align}
Let us introduce new notations reflecting these properties:
\begin{align}
    & \vec{r} = (x,y) := (\varepsilon X,\varepsilon Y),\quad \tau:= \varepsilon c_0 T,\quad z := Z,\\
    & \frac{\partial}{\partial x}  \sim \frac{\partial}{\partial y}  \sim \frac{\partial}{\partial \tau } \sim \frac{\partial}{\partial z},
\end{align}
The wave equation is rewritten in the form
\begin{align}
    & \left[- n^2(x,y,z) \frac{\partial ^2}{\partial \tau^2} +\frac{\partial ^2}{\partial x^2} + \frac{\partial ^2}{\partial y^2} + \varepsilon^{-2}\frac{\partial ^2}{\partial z^2} \right] u(\tau,x,y,z) = 0.
\end{align}
Let us start with an ansatz for the solution of such an equation in the following form:
\begin{align}
    & u (\tau,x,y,z) =  A(\tau,x,y)\psi(\tau,x,y,z)\exp{\left\{i\varepsilon^{-1}\phi(\tau,x,y)\right\}},
\end{align}
where
\begin{align}
    & A(\tau,x,y) := \sum_{j=0}^{\infty} (i\varepsilon)^j A^j(\tau,x,y).
\end{align}
In this work, we are going to focus on the propagation of the single mode without consideration of effects related to mode coupling. Substitution of the ansatz after the reduction of the exponential multiplier leads to the equation:
\begin{align}
    \notag& A \varepsilon^{-2}\bigg[\frac{\partial^2 \psi}{\partial z^2} + \left\{n^2 \left(\frac{\partial \phi}{\partial\tau}\right)^2  - \left(\vec{\nabla\phi}\right)^2\right\}\psi\bigg] + \\ \notag&i\varepsilon^{-1}\psi\bigg[A\left\{\Delta\phi -n^2 \frac{\partial^2\phi}{\partial\tau^2}\right\}  + 2 \left(\vec{\nabla} \phi,\vec{\nabla}A\right)-2n^2\frac{\partial\phi}{\partial\tau}\frac{\partial A}{\partial\tau}\bigg] +\\ \notag&+ 2i\varepsilon^{-1}A\bigg[{ -n^2\frac{\partial \phi}{\partial\tau}\frac{\partial \psi}{\partial\tau}}+ \left(\vec{\nabla} \phi,\vec{\nabla}\psi\right)\bigg] +  \\ &
   + \bigg[\Delta A \psi + A\Delta\psi - n^2{ A\frac{\partial^2\psi}{\partial\tau^2}}+ 2 \left(\vec{\nabla} A,\vec{\nabla}\psi\right) \bigg] = 0
\end{align}
The gradient $\vec{\nabla}$ and the Laplace operator $\Delta$ are the operators with respect to the variables $x$ and $y$. Let us define the instantaneous frequency and the instantaneous wave vector by \begin{align}
    w:= \frac{\partial \phi}{\partial \tau},\quad \vec{k}:= \frac{\partial \phi}{\partial \vec{r}} = \left(\frac{\partial \phi}{\partial x},\frac{\partial \phi}{\partial y}\right)^t=(k_x,k_y)^t.
\end{align}
Then the first two equations from decomposition by powers of $\varepsilon$ have the form:
\begin{align}
    \label{first order}& \bigg[\frac{\partial^2 \psi}{\partial z^2} +\left\{ n^2 w^2  - |\vec{k}|^2\right\}\psi\bigg]A^0 = 0,\\
    \notag& \bigg[A^0\left\{\Delta\phi -n^2\frac{\partial^2\phi}{\partial\tau^2}\right\}  +2 \left(\vec{\nabla} \phi,\vec{\nabla}A^0\right)-2n^2\frac{\partial\phi}{\partial\tau}\frac{\partial A^0}{\partial\tau}\bigg]\psi\,\,+\\ \label{second order}& + 2A^0\bigg[{ -n^2\frac{\partial \phi}{\partial\tau}\frac{\partial \psi}{\partial\tau}}+ \left(\vec{\nabla} \phi,\vec{\nabla}\psi\right)\bigg] = 0.
\end{align}
\subsection{Vertical modes}
Consider a family of Sturm-Liouville operators acting on the interval $[0,+\infty)$ and depending on $w,\vec{r}$ as parameters:
\begin{align}
    & L[w,\vec{r}]\psi:= \left[\frac{d^2}{d z^2} + n^2(\vec{r},z) w^2\right] \psi\\
    & \psi|_{z=0} = 0,\quad \psi|_{z = h(\vec{r})+} = \psi|_{z = h(\vec{r})-},\\ 
    & \frac{1}{\rho(\vec{r})^{+}} \frac{\partial \psi}{\partial z}\bigg|_{z = h(\vec{r})+}  = \frac{1}{\rho(\vec{r})^{-}} \frac{\partial\psi}{\partial z}\bigg|_{z = h(\vec{r})-},
\end{align}
and also $\psi$ satisfies the radiation conditions at infinity ($\psi\in L_2(\mathbb{R}_+)$). In general, such operators have finite positive discrete spectrum (the negative part is considered non-physical and usually is outside of the scope):
\begin{align}
&\lambda_0(w,\vec{r})>\lambda_1(w,\vec{r})>\dots>\lambda_{m-1}(w,\vec{r})>0,
\end{align}
We will assume that this part of the spectrum consists of simple eigenvalues (there is a one-to-one correspondence between eigenvalues  and eigenvectors). Generally speaking, the number of positive eigenvalues $m$ depends on the parameters $w,\vec{r}$. We will refer to the corresponding eigenfunctions $\psi_0,\dots,\psi_{m-1}$ as vertical modes. Due to the positivity of $\lambda$, we introduce the notation:
\begin{align}
& q_l (w,\vec{r}) := \sqrt{\lambda_l (w,\vec{r})} >0,\quad l =0 ,\dots, m-1.
\end{align}
The scalar product between the eigenfunctions is given by the following formula:
\begin{align}
    & \langle \psi_l,\psi_{l'}\rangle := \rho(\vec{r})^{+}\int_{0}^{h(\vec{r})}\psi_l\psi_{l'} dz + \rho(\vec{r})^{-}\int_{h(\vec{r})}^{\infty}\psi_l\psi_{l'} dz = \delta_{l l'}.
\end{align}
\subsection{Eikonal equation}
Let us consider in detail the equation \ref{first order}. Let $\psi = \psi_l$ be some fixed vertical mode. From this equation, we obtain (for arbitrary amplitude $A^0$) a nonlinear equation on the function $\phi$ (hereafter, we omit the index $l$ as all considerations are made for some fixed mode): 
\begin{align}
    &  q^2 \left(\vec{r},\frac{\partial \phi}{\partial \tau}\right) - |\vec{\nabla} \phi|^2   = 0.
\end{align}
We will refer to this equation as the eikonal equation. It can be solved using the standard method of characteristics. Let us introduce the Hamiltonian $\mathcal{H}$ by the following formula
\begin{align}
    & \mathcal{H} (w,\vec{k},\vec{r}) := |\vec{k}|^2 - q^2(w,\vec{r})\equiv 0.
\end{align}
Then the space-time rays in coordinates$(\tau,\vec{r})$ along with corresponding lines in momentum space ($w,\vec{k}$) are defined as solutions of the canonical Hamiltonian system \cite{babich1998space}:
\begin{align}
    & \frac{d}{d\xi}\left(\begin{array}{c}
         \hat{r}  \\
         \hat{k} 
    \end{array}\right) =\left(\begin{array}{cc}
           0  & \mathrm{I}_{3\times 3} \\
           -\mathrm{I}_{3\times 3}  & 0
        \end{array}\right) \left(\begin{array}{c}
         \hat{\nabla}_r \mathcal{H}  \\
         \hat{\nabla}_k \mathcal{H}
    \end{array}\right),
\end{align}
where $\xi$ is a one-dimensional parameter along the ray, which, generally speaking, does not have explicit physical meaning, and $\hat{\cdot}$ denotes objects referring to the three-dimensional space $(\tau,\vec{r})$ or corresponding momentum space $(w,\vec{k})$. The explicit form of such equations will be introduced later; for now, we just assume, that this canonical system is somehow solved. Let us introduce ray-centered coordinates $\xi,\mu,\nu$, such that
\begin{align}
    & \hat{r} = \hat{r}(\xi,\mu,\nu),\quad \hat{k} = \hat{k}(\xi,\mu,\nu),
\end{align}
and also a points in phase space along with ray coordinates
\begin{align}
    &\mathfrak{f} := \left(\begin{array}{l}
         \hat{r}   \\
         \hat{k} 
    \end{array}\right),&& \mathfrak{r}:= \left(\begin{array}{c}
          \xi   \\
         \mathfrak{r}^{\perp} 
    \end{array}\right)= \left(\begin{array}{c}
         \xi   \\
         (\mu,\nu)^t
    \end{array}\right). 
\end{align}
We should once again mention that coordinates $\mu,\nu$ ($\mathfrak{r}^{\perp}$) somehow characterize the initial parameters of the ray, and $\xi$ is the only parameter changing along the ray. Using notations 
\begin{align}
    & \nabla_{\mathfrak{f}}  :=\left(\hat{\nabla}_r,\hat{\nabla}_k\right)^t =\left(\frac{\partial}{\partial \tau},\vec{\nabla}_r,\frac{\partial}{\partial w},\vec{\nabla}_k\right)^t,
    && \nabla_{\mathfrak{r}}  :=\left(\frac{\partial}{\partial \xi},\frac{\partial}{\partial \mu},\frac{\partial}{\partial \nu}\right)^t,
\end{align}
we have a simple form of the Hamilton equations:
\begin{align}
    \label{ham syst}& \left\{\begin{array}{l}
         \frac{d}{d\xi} \mathfrak{f}(\xi,\mathfrak{r}^{\perp}_0) = \mathfrak{J} \nabla_{\mathfrak{f}} \mathcal{H}(\xi,\mathfrak{r}^{\perp}_0)  \\[1ex]
         \mathfrak{f}(\xi_0,\mathfrak{r}^{\perp}_0) =  \mathfrak{f}_0(\mathfrak{r}^{\perp}_0)
    \end{array}\right.
\end{align}
where $\mathfrak{J}$ is the standard symplectic matrix of the corresponding six-dimensional phase space used in previous equations.
\subsection{Transport equation}
Now we will consider the equation \ref{second order}. Let us multiply it by $A^0$ and rewrite it in a more convenient form (omitting index $0$):
\begin{align}
\notag& 
\bigg[A^2\left\{\Delta\phi -n^2\frac{\partial^2\phi}{\partial\tau^2}\right\}  +2 A\left(\vec{\nabla} \phi,\vec{\nabla}A\right)-2n^2A\frac{\partial\phi}{\partial\tau}\frac{\partial A}{\partial\tau}\bigg]\psi+\\ \label{trans eq}& + 2A^2\bigg[{ -n^2\frac{\partial \phi}{\partial\tau}\frac{\partial \psi}{\partial\tau}}+ \left(\vec{\nabla} \phi,\vec{\nabla}\psi\right)\bigg]= 0
\end{align}
It is not difficult to see that the dependence on the parameter $z$ is contained only in the functions $n$ and $\psi$. Consider the auxiliary equation \begin{align}
    & \langle\psi'',\psi\rangle + w^2 \langle n^2 \psi,\psi\rangle = q^2 \langle\psi,\psi\rangle,
\end{align}
which is obtained by scalar multiplication of the equation $L[w,\vec{r}]\psi = q^2 \psi$ by the same eigenfunction $\psi$. It shows that
\begin{align}
    & \langle n^2\psi,\psi\rangle = \frac{1}{w^2} (q^2 + \langle\psi',\psi'\rangle)\approx \frac{q}{w} \frac{\partial q}{\partial w}.
\end{align}
Let us also compute expressions for the other scalar products that we will need:
\begin{align}
    & \langle (\vec{\nabla}\phi,\vec{\nabla}\psi),\psi\rangle = \frac{1}{2} (\vec{\nabla} \phi,\vec{\nabla}\langle\psi,\psi\rangle) = 0,\\
        & \langle n^2\frac{\partial \psi}{\partial \tau},\psi\rangle = \frac{1}{2}\frac{\partial}{\partial\tau} \langle n^2\psi,\psi\rangle\approx \frac{1}{2}\frac{\partial}{\partial\tau}\left[\frac{q}{w}\frac{\partial q}{\partial w}\right].
\end{align}
Then, taking the scalar product of the transport equation \ref{trans eq} with the function $\psi$, we have
\begin{align}
&  \frac{\partial}{\partial \tau} \left[-A^2 w\langle n^2\psi,\psi\rangle\right]+\vec{\nabla}\cdot \left[A^2\vec{k}\right]\langle\psi,\psi\rangle = 0
\end{align}
This equation can be written in divergence form:
\begin{align}
    & \hat{\mathrm{div}} \left(\frac{q}{v}A^2\hat{\kappa}'\right) = 0,&& \hat{\kappa}'=\left(-\frac{v}{q}w\langle n^2\psi,\psi\rangle,\frac{v}{q}\vec{k}\right)^t.
\end{align}
Using the approximation for $\langle n^2\psi,\psi\rangle $, we obtain a more convenient form for the transport equation of the principal component of the amplitude:
\begin{align}
    \label{transport approx}& \hat{\mathrm{div}} \left(\frac{q}{v}A^2\hat{\kappa}\right) = 0,&& \hat{\kappa}=\left(1,\frac{v}{q}\vec{k}\right)^t.
\end{align}
\subsection{First variation equations}
As we are interested in some details of wave propagation besides the rays themselves we should use variational methods for Hamilton systems \cite{farra1993ray,farra1989ray}. Let us take the variation of the Hamilton system \ref{ham syst}:
\begin{align}
    \frac{d}{d\xi}\delta\mathfrak{f}=\delta\left(\frac{d}{d\xi} \mathfrak{f}\right) =   \mathfrak{J} \delta\left(\nabla_{\mathfrak{f}} \mathcal{H}\right)= \mathfrak{J} \nabla_{\mathfrak{f}}\nabla_{\mathfrak{f}}^t \mathcal{H} \delta\mathfrak{f} .
\end{align}
This means that for small variations of the original ray parameters (along some  \textit{reference} ray with initial parameters $\mathfrak{r}^{\perp}= \mathfrak{r}^{\perp}_0$):
\begin{align}
     &\delta\mathfrak{f}(\xi,\mathfrak{r}^{\perp},\mathfrak{r}^{\perp}_0) :=  \left(\begin{array}{c}
            \delta\hat{r}(\xi,\mathfrak{r}^{\perp},\mathfrak{r}^{\perp}_0)\\
          \delta\hat{k}(\xi,\mathfrak{r}^{\perp},\mathfrak{r}^{\perp}_0)
     \end{array}\right)= \left(\begin{array}{c}
            \hat{r}(\xi,\mathfrak{r}^{\perp}) -  \hat{r}(\xi,\mathfrak{r}^{\perp}_0)\\
          \hat{k}(\xi,\mathfrak{r}^{\perp}) -  \hat{k}(\xi,\mathfrak{r}^{\perp}_0)
     \end{array} \right)
\end{align}
we have the system\footnote{In all equations, the matrix $\nabla_{\mathfrak{f}}\nabla_{\mathfrak{f}}^t \mathcal{H}$ is the Hessian matrix $\mathrm{H}_{\mathfrak{f}}(\mathcal{H})$ of the Hamiltonian, but the explicit form using the $\nabla$-operator appears to be more useful later.} 
\begin{align}
    \label{first var syst}& \left\{\begin{array}{l}
         \frac{d}{d\xi}\delta\mathfrak{f}(\xi,\mathfrak{r}^{\perp},\mathfrak{r}^{\perp}_0) = \mathfrak{J} \nabla_{\mathfrak{f}}\nabla_{\mathfrak{f}}^t\mathcal{H}(\xi,\mathfrak{r}^{\perp}_0)\delta\mathfrak{f}(\xi,\mathfrak{r}^{\perp},\mathfrak{r}^{\perp}_0) \\[1ex]
         \delta\mathfrak{f}(\xi_0,\mathfrak{r}^{\perp},\mathfrak{r}^{\perp}_0)  = \delta\mathfrak{f}_0(\mathfrak{r}^{\perp},\mathfrak{r}^{\perp}_0)
    \end{array}\right. .
\end{align}
It appears to be convenient to study this system via the propagator matrix (also known as the monodromy matrix) $\mathcal{P}(\xi,\xi_0, \mathfrak{r}^{\perp}_0)$ defined by the following equations\footnote{Here and below, we omit $\mathfrak{r}^{\perp}_0$ for simplicity unless it is necessary.}:
\begin{align}
    \label{propagator syst}& \left\{\begin{array}{l}
         \frac{d}{d\xi}\mathcal{P}(\xi,\xi_0) = \mathfrak{J} \nabla_{\mathfrak{f}}\nabla_{\mathfrak{f}}^t\mathcal{H}(\xi)\mathcal{P}(\xi,\xi_0)\\[1ex]
         \mathcal{P}(\xi_0,\xi_0) = \mathrm{I}_{6\times 6}
    \end{array}\right..
\end{align}
Any propagator matrix defined by the system above is symplectic:
\begin{align}
    & \mathcal{P}^t \mathfrak{J}\mathcal{P} =\mathfrak{J}, \quad \mathcal{P}^{-1} = \mathfrak{J}^{t}\mathcal{P}^t\mathfrak{J}, \quad \det\,\mathcal{P}(\xi) = \det\,\mathrm{I}_{6\times 6} =  1.
\end{align}
Such a matrix can also be expressed in the direct form in terms of the ordered exponential (also known as the $\mathcal{T}$-exponential or path exponential):
\begin{align}
    & \mathcal{P}(\xi,\xi_0) = \mathcal{T}\exp\left\{\int_{\xi_0}^{\xi}\mathfrak{J} \nabla_{\mathfrak{f}}\nabla_{\mathfrak{f}}^t\mathcal{H}(\xi')d\xi'\right\},
\end{align}
which has the obvious consequence (for the same values $\mathfrak{r}^{\perp}_0$):
\begin{align}
    & \mathcal{P}(\xi,\xi')\mathcal{P}(\xi',\xi_0) = \mathcal{P}(\xi,\xi_0),\quad \xi_0\le\xi'\le\xi.
\end{align}
From the definition of the propagator, we have the formula for small variations:
\begin{align}
    & \delta\mathfrak{f}(\xi)= \mathcal{P}(\xi,\xi_0)\delta\mathfrak{f}_0.
\end{align}
Let us introduce Jacobi matrices linking different coordinates:
\begin{align}
     \mathcal{J}_{\mathfrak{f}} (\mathfrak{r}) &:= \frac{\partial \mathfrak{f}}{\partial \mathfrak{r}} =\frac{\partial (\tau,x,y,w,k_x,k_y)}{\partial (\xi,\mu,\nu)} = \left(\begin{array}{ccc}
        \frac{\partial \tau}{\partial \xi} & \frac{\partial \tau}{\partial \mu} &\frac{\partial \tau}{\partial \nu}\\[1ex]
        \frac{\partial \vec{r}}{\partial \xi} & \frac{\partial \vec{r}}{\partial \mu} &\frac{\partial \vec{r}}{\partial \nu}\\[1ex]
        \frac{\partial w}{\partial \xi} & \frac{\partial w}{\partial \mu} &\frac{\partial w}{\partial \nu}\\[1ex]
        \frac{\partial \vec{k}}{\partial \xi} & \frac{\partial \vec{k}}{\partial \mu} &\frac{\partial \vec{k}}{\partial \nu}
    \end{array}\right),\\
     \mathcal{J}_{\mathfrak{f}} (\mathfrak{r}) &= \left(\mathcal{J}_{\mathfrak{f}}^{\xi},\mathcal{J}_{\mathfrak{f}}^{\mu},\mathcal{J}_{\mathfrak{f}}^{\nu}\right).
\end{align}
From the general theory of Hamiltonian systems, we know that the following statement holds for any point in phase space $\mathfrak{f} = \mathfrak{f}(\mathfrak{r})$:
\begin{align}
    & \mathrm{rank}\,\, \mathcal{J}_{\mathfrak{f}}(\mathfrak{r}) = 3
\end{align}
if such a condition holds for the initial data:
\begin{align}
    & \mathrm{rank}\,\, \mathcal{J}_{\mathfrak{f}}(\xi_0,\mathfrak{r}^{\perp}) = 3.
\end{align}
We assume that the initial data are such that the condition above is satisfied. This means that at each point there are \textit{at least} 3 coordinates in phase space with a non-degenerate Jacobi transformation matrix. This arises from the fact that the Hamiltonian defines a 3-dimensional  Lagrange manifold in phase space, which is locally projectable onto some 3-dimensional plane in 6-dimensional space\footnote{For example, it is also useful for the construction of global asymptotics of solutions using Maslov's canonical operator - see \cite{dobrokhotov2017new,maslov2001semi,petrov2019application}.}. It is easy to show, that the full Jacobi matrix satisfies the equation
\begin{align}
    & \mathcal{J}_{\mathfrak{f}}(\xi,\mathfrak{r}^{\perp}_0) = \mathcal{P}(\xi,\xi_0,\mathfrak{r}^{\perp}_0)\mathcal{J}_{\mathfrak{f}}(\xi_0,\mathfrak{r}^{\perp}_0).
\end{align}
Due to the fact, that Jacobi matrices have full rank (linearly independent columns), we can define Moore–Penrose inverse matrices with the formula
\begin{align}
    & \mathcal{J}_{\mathfrak{f}}^{+} := \left(\mathcal{J}_{\mathfrak{f}}^t\mathcal{J}_{\mathfrak{f}}\right)^{-1}\mathcal{J}_{\mathfrak{f}}^t,
\end{align}
while such $\mathcal{J}_{\mathfrak{f}}^{+}$ appears to be a left inverse for the matrix $\mathcal{J}_{\mathfrak{f}}$:
\begin{align}
    & \mathcal{J}_{\mathfrak{f}}^{+} \mathcal{J}_{\mathfrak{f}} = \mathrm{I}_{3\times 3}.
\end{align}
Considering small variations of phase space and ray coordinates, we have the relations:
\begin{align}
    & \delta \mathfrak{f} = \mathcal{J}_{\mathfrak{f}} \delta \mathfrak{r}, && \delta \mathfrak{r} = \mathcal{J}_{\mathfrak{f}}^{+}\delta \mathfrak{f}.  
\end{align}
Then for some function $\mathcal{F}$ of phase variables $\mathfrak{f}$ depending on ray coordinates $\mathfrak{r}$, we have (neglecting the smaller terms)
\begin{align}
&\delta \mathcal{F}=  \delta \mathfrak{f}^t\nabla_{\mathfrak{f}}\mathcal{F}   = \delta \mathfrak{r}^t\nabla_{\mathfrak{r}}\mathcal{F}.
\end{align}
Using relations between variations of coordinates, we have weak relations between gradients:
\begin{align}
    & \delta \mathfrak{f}^t \left(\nabla_{\mathfrak{f}}\mathcal{F} - (\mathcal{J}_{\mathfrak{f}}^{+})^t\nabla_{\mathfrak{r}}\mathcal{F}   \right) = \delta \mathfrak{r}^t \left(\nabla_{\mathfrak{r}}\mathcal{F} - \mathcal{J}_{\mathfrak{f}}^t\nabla_{\mathfrak{f}}\mathcal{F}   \right) = 0.
\end{align}
As a result, we have 
\begin{align}
     \nabla_{\mathfrak{r}}\mathcal{F} &= \mathcal{J}_{\mathfrak{f}}^t\nabla_{\mathfrak{f}}F,&
     \nabla_{\mathfrak{f}}\mathcal{F} &=(\mathcal{J}_{\mathfrak{f}}^{+})^t\nabla_{\mathfrak{r}}\mathcal{F}.
\end{align}
Finally, in terms of the propagator matrix and the initial value for the Jacobi matrix, we have
\begin{align}
    & \mathcal{J}_{\mathfrak{f}}^t = \mathcal{J}_{\mathfrak{f}_0}^t \mathcal{P}^t, && (\mathcal{J}_{\mathfrak{f}}^{+})^t = \mathcal{P} \mathcal{J}_{\mathfrak{f}_0}\left(\mathcal{J}_{\mathfrak{f}_0}^t\mathcal{P}^t\mathcal{P}\mathcal{J}_{\mathfrak{f}_0}\right)^{-1}.   
\end{align}
We can easily see that the derivatives $\nabla_{\mathfrak{r}} \mathcal{F}$ depend on the choice of ray coordinates, but ${\nabla}_{\mathfrak{f}} \mathcal{F}$ is invariant under such changes. In the contrast to other gradients, $\nabla_{\mathfrak{r}} \mathcal{F}$ can usually be computed along the reference ray. With this in mind, we will call  ${\nabla}_{\mathfrak{f}} \mathcal{F}$  \textit{observable} and $\nabla_{\mathfrak{r}} \mathcal{F}$ \textit{computative} gradients, respectively. 
\subsection{Second variation equations}
Even though first variation equations allow us to compute the dynamics of Jacobian matrices along the ray, we still can't compute their variations due to small changes in initial data other than directly. To solve this problem, we can use second variation equations. Let us start with the definition of the variation of the propagator:
\begin{align}
    & \delta \mathcal{P}(\xi,\xi_0,\mathfrak{r}^{\perp}, \mathfrak{r}_0^{\perp}) : =  \mathcal{P}(\xi,\xi_0,\mathfrak{r}^{\perp}) -  \mathcal{P}(\xi,\xi_0, \mathfrak{r}_0^{\perp}). 
\end{align}
For such variation, we have the system  
\begin{align}
    \label{prop var syst}& \left\{\begin{array}{l}
         \frac{d}{d\xi}\delta\mathcal{P} = \mathfrak{J} \nabla_{\mathfrak{f}}\nabla_{\mathfrak{f}}^t\mathcal{H}\delta\mathcal{P} + \mathfrak{J} \nabla_{\mathfrak{f}}^3\mathcal{H}[\mathcal{P}, \mathcal{P}] \mathcal{J}_{\mathfrak{f_0}}\delta \mathfrak{r},  \\[1ex]
         \delta\mathcal{P}(\xi_0,\mathfrak{r}^{\perp},\mathfrak{r}^{\perp}_0)  = 0_{6\times 6}
    \end{array}\right. .
\end{align}
Here $\nabla_{\mathfrak{f}}^3\mathcal{H}$ is a rank-3 tensor acting on the bilinear form $[\mathcal{P}, \mathcal{P}]$ by the rule
\begin{align}
    & (\nabla_{\mathfrak{f}}^3\mathcal{H}[\mathcal{P}, \mathcal{P}])_{ijk} = \left(\nabla_{\mathfrak{f}}^3\mathcal{H}\right)_{ilm} \mathcal{P}_{lj}\mathcal{P}_{mk}
\end{align}
and then that contracts with $\mathcal{J}_{\mathfrak{f}}\delta \mathfrak{r}_0$
\begin{align}
    (\nabla_{\mathfrak{f}}^3\mathcal{H}[\mathcal{P}, \mathcal{P}]\mathcal{J}_{\mathfrak{f}}\delta \mathfrak{r}_0)_{ij} = (\nabla_{\mathfrak{f}}^3\mathcal{H}[\mathcal{P}, \mathcal{P}])_{ijk} (\mathcal{J}_{\mathfrak{f_0}})_{kl}(\delta \mathfrak{r})_{l}.
\end{align}
Due to the fact that the propagator $\mathcal{P}$ is a fundamental solution of such 
a system, by definition, we have the explicit solution 
\begin{align}
    & \delta\mathcal{P}(\xi,\delta\mathfrak{r}_0) =   \mathcal{P}(\xi) \int_{\xi_0}^{\xi} \mathcal{P}^{-1}(\xi') \mathfrak{J}\nabla_{\mathfrak{f}}^3\mathcal{H}[\mathcal{P}, \mathcal{P}] \mathcal{J}_{\mathfrak{f_0}}d\xi' \delta \mathfrak{r}
\end{align}
The same result appears if we introduce the rank-3 propagation tensor $\Phi$
\begin{align}
    & \Phi(\xi) =   \mathcal{P}(\xi) \int_{\xi_0}^{\xi} \mathcal{P}^{-1}(\xi') \mathfrak{J}\nabla_{\mathfrak{f}}^3\mathcal{H}[\mathcal{P}, \mathcal{P}] d\xi'
\end{align}
as a solution of the system
\begin{align}
    & \label{second prop syst}& \left\{\begin{array}{l}
         \frac{d}{d\xi}\Phi = \mathfrak{J} \nabla_{\mathfrak{f}}\nabla_{\mathfrak{f}}^t\mathcal{H}\Phi + \mathfrak{J} \nabla_{\mathfrak{f}}^3\mathcal{H}[\mathcal{P}, \mathcal{P}],   \\[1ex]
         \Phi(\xi_0)  = 0_{6\times 6\times 6}
    \end{array}\right. .
\end{align}
Then derivatives of the propagator with respect to ray coordinates can be expressed in the form
\begin{align}
    &  \nabla_{\mathfrak{r}}\mathcal{P} = \Phi [\mathcal{J}_{\mathfrak{f}_0},\cdot],&&  (\nabla_{\mathfrak{r}}\mathcal{P})_{ijk} = \Phi_{ijl} (\mathcal{J}_{\mathfrak{f}_0})_{lk}.
\end{align}
As a result, we have all derivatives of the propagator $\mathcal{P}$ expressed in terms of the solution of some differential equation along the reference ray. 
\subsection{Change of variables}
It was mentioned before that the variable $\xi$ along the ray does not have an explicit physical meaning. This raises the question: does anything in the considerations above change with the transition from $\xi$ to some $\zeta$ (also a variable along the ray)? The answer is affirmative. Let the variable $\zeta$ be defined via the relation
\begin{align}
    & d \xi :=\sigma(\mathfrak{f}(\zeta)) d\zeta.
\end{align}
Then we have the Hamilton equation in the form
\begin{align}
    &         \frac{d}{d\zeta} \mathfrak{f}(\zeta) = \mathfrak{J} \sigma(\zeta)\nabla_{\mathfrak{f}} \mathcal{H}(\zeta),
\end{align}
so the zeroth order of the equations differs only by the multiplicative factor $\sigma$.  However, for the first and second order equations (let us demonstrate it just for general propagator equations), we observe a significant change:
\begin{align}
    & \frac{d}{d\zeta} \mathcal{P}(\zeta) = \mathfrak{J} \left\{\nabla_{\mathfrak{f}}\sigma\nabla_{\mathfrak{f}}^t\right\} \mathcal{H} \mathcal{P} =  \mathfrak{J} \sigma\nabla_{\mathfrak{f}}\nabla_{\mathfrak{f}}^t \mathcal{H} \mathcal{P} + \mathfrak{J} (\nabla_{\mathfrak{f}}\sigma)(\nabla_{\mathfrak{f}}^t \mathcal{H}) \mathcal{P},\\
    & \frac{d}{d\zeta}\Phi(\zeta) = \mathfrak{J} \left\{\nabla_{\mathfrak{f}}\sigma\nabla_{\mathfrak{f}}^t\right\}\mathcal{H}\Phi + \mathfrak{J} \left\{\nabla_{\mathfrak{f}}^2\sigma\nabla_{\mathfrak{f}}^t\right\}\mathcal{H}[\mathcal{P}, \mathcal{P}].
\end{align}
This change is necessary due to the fact that different Hamiltonians can define the same rays while simultaneously having significantly different first and second variations, as well as different natural variables. Thus, the modified equations for these variations allow us to reach the same result, invariant to such changes.
\subsection{Explicit forms of equations}
Let us start with the Hamilton system \ref{ham syst}. Substituting the explicit form of the Hamiltonian, we have 
\begin{align}
    &      \frac{d}{d\xi}\left(\tau,\vec{r},w,\vec{k}\right)^t = 2 q\left(-\frac{\partial q}{\partial w},\frac{\vec{k}}{q},0,\vec{\nabla}q\right)^t.
\end{align}
It is easy to see, that there are two convenient choices of variables along the ray:
\begin{align}
    & d s:= 2 q d\xi,&& d\tau : = \frac{d s}{v} ,\quad v:= - \left(\frac{\partial q}{\partial w}\right)^{-1}.
\end{align}
Here, $s$ corresponds to the length of the spatial part (projection) of the ray, and $\tau$ represents the time of propagation along the ray with group speed $v$. Using $\tau$ as a variable, we have
\begin{align}
    \label{ham explicit}&      \frac{d}{d\tau}\left(\tau,\vec{r},w,\vec{k}\right)^t = \left(1,\frac{v}{q}\vec{k},0,v \vec{\nabla}q\right)^t.
\end{align}
For the variable $\tau$, we have the factor $\sigma$ as follows:
\begin{align}
    & \sigma(\tau) := \frac{v(\tau)}{2q(\tau)}.
\end{align}
\subsection{Transport equation and phase change along ray}
Let us now take a closer look at the transport equation:
\begin{align}
& \hat{\mathrm{div}} \left(\frac{q}{v}A^2\hat{\kappa}\right) = 0,
\end{align}
Considering an infinitesimal ray tube along the space-time ray (in the direction of the three-dimensional vector $\hat{\kappa}$), we obtain the invariance of the expression
\begin{align}
    &  A^2 \frac{q}{v}|\hat{\kappa}| d \mathcal{S} = A^2 \frac{q}{v}\sqrt{1+v^2} d \mathcal{S}= A^2 q\sqrt{1+v^{-2}} d \mathcal{S} \equiv \mathrm{const},
\end{align}
where $d \mathcal{S}$ is the cross-sectional element of this tube. Hence, we find that, for the amplitude along the ray, the following formula is valid:
\begin{equation}\label{a(s) first}
    A(\tau) = A(0) \times \sqrt{\frac{ g(0)d\mathcal{S}(0)}{g(\tau)d\mathcal{S}(\tau)}}, \quad  g(\tau):=q\sqrt{1+v^{-2}}.
\end{equation}
An explicit expression can be obtained for the ratio of the cross-section elements of a ray tube in terms of determinants of Jacobi matrices:
\begin{align}
& \frac{d\mathcal{S}(0)}{d\mathcal{S}(\tau)} = \frac{\mathfrak{D}(0)}{\mathfrak{D}(\tau)}, \quad \mathfrak{D}(\tau):= \left|\frac{\partial (\tau,\vec{r})}{\partial \mathfrak{r}}\right| = |\det \mathcal{J}_{\hat{r}}| .
\end{align}               
Then the formula \ref{a(s) first} can be written in the following form:
\begin{equation}\label{a(s) second}
    A(\tau,\mathfrak{r}^{\perp}) = A(0,\mathfrak{r}^{\perp}) \times \sqrt{\frac{ g(0,\mathfrak{r}^{\perp})\mathfrak{D}(0,\mathfrak{r}^{\perp})}{g(\tau,\mathfrak{r}^{\perp})\mathfrak{D}(\tau,\mathfrak{r}^{\perp})}}.
\end{equation}
Now let us consider the change of the phase $\phi$ along the ray. From the definition of the wave vector $\hat{k}$, it follows that
\begin{equation}
    d \phi  = w d\tau + (\vec{k},d\vec{r}) = (w + vq)d\tau.
\end{equation}
Then 
\begin{equation}
    \phi(\tau,\mathfrak{r}^{\perp}) = \phi_0(\mathfrak{r}^{\perp}) + w\tau + \int_{0}^{\tau}q(\tau',\mathfrak{r}^{\perp})v(\tau',\mathfrak{r}^{\perp})d\tau'.
\end{equation}
\section{Fronts and observations}
Let us once again consider some arbitrary function $\mathcal{F}$ of coordinates $\mathfrak{f}$. 
\subsection{General fronts}
We will say that such a function has 
\begin{itemize}
    \item a space front at the point $\mathfrak{f}'$ if
\begin{align}
    & \vec{\nabla}_{r} \mathcal{F}\big|_{\mathfrak{f}=\mathfrak{f}'} = \left(\frac{\partial \mathcal{F}}{\partial x}, \frac{\partial\mathcal{F}}{\partial y}\right)^t\bigg|_{\mathfrak{f}=\mathfrak{f}'}\neq 0;
\end{align}
    \item a space-time front at the point $\mathfrak{f}'$ if
\begin{align}
    & \hat{\nabla}_{r} \mathcal{F}\big|_{\mathfrak{f}=\mathfrak{f}'}  = \left(\frac{\partial \mathcal{F}}{\partial \tau},\frac{\partial \mathcal{F}}{\partial x}, \frac{\partial \mathcal{F}}{\partial y}\right)^t\bigg|_{\mathfrak{f}=\mathfrak{f}'}\neq 0.
\end{align}
\end{itemize}
Such gradients for fronts can be computed as follows:
\begin{align}
     &\vec{\nabla}_{r} \mathcal{F}\big|_{\mathfrak{f} = \mathfrak{f}'} = \left[(\mathcal{J}_{\mathfrak{f}}^{+})^t\nabla_{\mathfrak{r}}\mathcal{F}\right]_{\vec{r}}, && 
     \hat{\nabla}_{r} \mathcal{F}\big|_{\mathfrak{f}=\mathfrak{f}'} = \left[(\mathcal{J}_{\mathfrak{f}}^{+})^t\nabla_{\mathfrak{r}}\mathcal{F}\right]_{\hat{r}}.
\end{align}
As it can be easily seen from the definition, the existence of the front (any one of those described above) on the initial surface guarantees such existence in some neighborhood of this surface. We would be interested in several examples of such fronts.
\subsection{Time and distance}
Let us consider the simplest examples of the function $\mathcal{F}$. 
\begin{itemize}
    \item For $\mathcal{F} = \tau$, we have a ray gradient
    \begin{align}
        & \nabla_{\mathfrak{r}}{\tau} = \left(1, 0,0\right)^t. 
    \end{align}
    Then related fronts are the result of propagating the initial space-time manifolds to the time $\tau$.
    \item For $\mathcal{F} = s(\tau)$, we have a ray gradient
    \begin{align}
        & \nabla_{\mathfrak{r}}{s} = \left(v, \int_{0}^{\tau} (\mathcal{J}_{\mathfrak{f}}^{\mu})^t{\nabla_{\mathfrak{f}}v}d\tau',\int_{0}^{\tau} (\mathcal{J}_{\mathfrak{f}}^{\nu})^t{\nabla_{\mathfrak{f}}v}d\tau'\right)^t. 
    \end{align}
    Then related fronts describe the boundary of the $s$-neighborhood of the initial space-time manifold.
\end{itemize}

\subsection{Phase}
Let us consider phase as function $\mathcal{F}$. As it was mentioned before, phase along the ray is described as follows: 
\begin{equation}
    \phi(\tau,\mathfrak{r}^{\perp}) = \phi_0(\mathfrak{r}^{\perp}) + w\tau + \int_{0}^{\tau}q(\tau',\mathfrak{r}^{\perp})v(\tau',\mathfrak{r}^{\perp})d\tau'.
\end{equation}
Then, for the derivatives of the phase, we have the expressions
\begin{align}
    & \frac{\partial \phi}{\partial \tau} = w + v q;\\
    & \frac{\partial \phi}{\partial \mu} = \frac{\partial \phi_0}{\partial \mu} + \frac{\partial w}{\partial \mu}\tau + \int_{0}^{\tau}(\mathcal{J}_{\mathfrak{f}}^{\mu})^t\left\{{v\nabla_{\mathfrak{f}} \,q} + q{\nabla_{\mathfrak{f}}\,v}\right\}d\tau',\\
        & \frac{\partial \phi}{\partial \nu} = \frac{\partial \phi_0}{\partial \nu} + \frac{\partial w}{\partial \nu}\tau + \int_{0}^{\tau}(\mathcal{J}_{\mathfrak{f}}^{\nu})^t\left\{{v\nabla_{\mathfrak{f}} \,q} + q{\nabla_{\mathfrak{f}}\,v}\right\}d\tau'.
\end{align}
Here function $\phi_0$ is defined (up to a constant) by the derivatives relations at the initial surface:
\begin{align}
    & \nabla_{\mathfrak{r}}\phi_0 := \mathcal{J}^t_{\mathfrak{f}_0}\nabla_{\mathfrak{f}}\phi_0 = \mathcal{J}^t_{\mathfrak{f}_0} (w_0,\vec{k}_0,0,\vec{0})^t,\\
    & \phi_0(\mu,\nu) = \int_{\mu_0}^{\mu} (\mathcal{J}_{\mathfrak{f}_0}^{\mu})^t\nabla_{\mathfrak{f}}\phi_0 d\mu' + \int_{\nu_0}^{\nu} (\mathcal{J}_{\mathfrak{f}_0}^{\nu})^t\nabla_{\mathfrak{f}}\phi_0 d\nu'.
\end{align}
Using these results, we can compute gradients in both space and space-time coordinates.
\subsection{Amplitude}
Let us consider amplitude (its principal part) $A$ as a function $\mathcal{F}$. For $A$ along the ray, we have the formula
\begin{align}
 &A(\tau,\mathfrak{r}^{\perp}) = A(0,\mathfrak{r}^{\perp}) \times \sqrt{\frac{ g(0,\mathfrak{r}^{\perp})\mathfrak{D}(0,\mathfrak{r}^{\perp})}{g(\tau,\mathfrak{r}^{\perp})\mathfrak{D}(\tau,\mathfrak{r}^{\perp})}}= A_0\sqrt{\frac{ g_0 \mathfrak{D}_0}{g \mathfrak{D}}},\\ &g=q\sqrt{1+{v}^{-2}},\quad \mathfrak{D} = |\det \mathcal{J}_{\hat{r}}|
\end{align}
The fact that amplitude propagates along the space-time ray originates from the comparison of the approximation of the transport equation \ref{transport approx} with the explicit form of Hamilton equations \ref{ham explicit}.  There is a convenient way to write the formula for the gradient in the form:
\begin{align}
      \nabla_{\mathfrak{\tau}} A &= A\big(A_0^{-1}\nabla_{\mathfrak{\tau}} A_0 + \frac{1}{2}\left\{g_0^{-1} \nabla_{\mathfrak{\tau}} g_0  - g^{-1} \nabla_{\mathfrak{\tau}} g\right\} +\\ & +\frac{1}{2}\left\{\mathfrak{D}_0^{-1} \nabla_{\mathfrak{\tau}} \mathfrak{D}_0-\mathfrak{D}^{-1} \nabla_{\mathfrak{\tau}} \mathfrak{D}\right\}\big).
\end{align}
For the derivatives of $\mathfrak{D}$, we would use a general relation (assuming that the determinant does not change sign along the ray)
\begin{align}
     &\mathfrak{D}^{-1} \nabla_{\mathfrak{r}} \mathfrak{D} =\left(\mathrm{tr} \left(\mathcal{J}_{\hat{r}}^{-1} \frac{\partial \mathcal{J}_{\hat{r}}}{\partial \tau}\right),\mathrm{tr} \left(\mathcal{J}_{\hat{r}}^{-1} \frac{\partial \mathcal{J}_{\hat{r}}}{\partial \mu}\right),\mathrm{tr} \left(\mathcal{J}_{\hat{r}}^{-1} \frac{\partial \mathcal{J}_{\hat{r}}}{\partial \nu}\right) \right)^t 
\end{align}
Thus, we need to compute the derivatives of Jacobi matrices. This can be done via equations for the second variation:
\begin{align}
    & \nabla_{\mathfrak{r}}\mathcal{J}_{\mathfrak{f}} = \nabla_{\mathfrak{r}}\mathcal{P}  \mathcal{J}_{\mathfrak{f}_0} + \mathcal{P}\nabla_{\mathfrak{r}}  \mathcal{J}_{\mathfrak{f}_0}.
\end{align}
Here, $\mathcal{J}_{\mathfrak{f}_0}$ and its derivatives play the role of initial data while, $\mathcal{P}$ with derivatives is computed along the ray. For the function $g$ we have a simple relation
\begin{align}
    & g^{-1}\nabla_{\mathfrak{r}} g = q^{-1}{\nabla_{\mathfrak{r}}q}- \frac{1}{v^2 + 1}v^{-1}{\nabla_{\mathfrak{r}}v}.
\end{align}
Amplitude $A_0$ is initially defined as a function in phase space coordinates $\mathfrak{f}$. Then the gradient of amplitude in ray coordinates can be computed via
\begin{align}
    & \nabla_{\mathfrak{r}} A_0 = \mathcal{J}_{\mathfrak{f}_0}^t\nabla_{\mathfrak{f}} A_0.
\end{align}
The equations above allow us to compute the derivative of amplitude $A$ at any arbitrary point on the ray, thus providing space and space-time gradients. 
\subsection{Observations}
It was shown previously that it is possible to compute space-time gradients of amplitude and phase at any given space-time point. All this allows us to describe the dynamics of the \textit{observable} field in the neighborhood of some fixed point. Since we can obviously observe only in space-time (not in momentum space), let us begin with a single-parameter observation line in space-time $\hat{r} = \hat{r}(\rho),\,\, \hat{r}(0) = \hat{r}_0 $ with some ray coordinates $\mathfrak{r}_0$ corresponding to  $\hat{r}_0$ (they also simultaneously define the starting momentum coordinates $\hat{k}_0$). Then the phase-space observation line $\mathfrak{f}$ (assuming we observe outside of caustics) and ray-coordinates system can be defined as follows:
\begin{align}
    \frac{d \mathfrak{f}}{d\rho} &= \mathcal{J}_{\mathfrak{f}}\mathcal{J}_{\hat{r}}^{-1}\frac{d \hat{r}}{d \rho} = \left(\frac{d \hat{r}}{d \rho}, \frac{d \hat{k}}{d \rho}\right)^t, &&\frac{d \mathfrak{r}}{d\rho} = \mathcal{J}_{\hat{r}}^{-1}\frac{d \hat{r}}{d \rho}. 
\end{align}
We are not going to focus on the analytical solutions of the equations above and will just assume that they are somehow solved (in applications they can be solved numerically with appropriate accuracy). Then, having all coordinates as functions of the parameter $\rho$ we have characteristics of the field (observable frequency and wave vector, amplitude space-time gradient) as functions of $\rho$. This allows us to understand the distribution of the acoustic field in the neighborhood of the observation line as a result of the propagation of the pulse as a whole \textit{automatically}, considering all interference effects of close (in space-time) rays. For example, we observe the effect of the difference between the observable frequency and the emitted one: even though the frequency $w$ \textit{as a parameter} doesn't change along the ray (due to the time-independence of the medium), the observable one can change significantly due to the compression/decompression of the pulse as a whole.

It is also important also to mention that this construction obviously does not take into account the multi-beam case when we have a set of rays with sufficiently different ray coordinates reaching the same space-time point. In such a case, we have to perform the regular summation of the set of fields, which is still much simpler than using the regular methods of consideration for all arriving space rays with different frequencies. 
\section{Possible applications and open questions}
\subsection{Observation at the point}
One of the important applications of the method described in our article is the prediction or explanation of the observed signal with initial frequency modulation at some fixed space point $\vec{r}$. In such a case, we can take the following as our observation line
\begin{align}
    & \hat{r}(\rho) := (\tau + \rho,\vec{r}).
\end{align}
Then we can, for example, compute time compression:
\begin{align}
     \delta\tau(\rho,\rho')&:=\tau(\rho) - \tau(\rho') = \delta\rho,\\ \delta\tau_0(\rho,\rho')&:=\tau_0(\rho) - \tau_0(\rho'),
\end{align}
as well as compare emitted and observable frequency modulations
\begin{align}
& w_0(\rho) :=  w_0(\mu(\rho),\nu(\rho)), && w(\rho) := \frac{\partial\phi}{\partial \tau} (\rho).
\end{align}
Another effect would be the computation and dynamics of the angle between amplitude and phase space fronts, as they can be observed in the experiments using several points of observation: the phase front based on the space and time dynamics of the appearing  signal and the amplitude front based on the intensity of signals. Such an angle, obviously, can be computed as an angle between observable gradients of amplitude and phase using the approach described above.  
\subsection{Outlook}
While our method provides an extensive description of the propagation of the nearly arbitrarily modulated signals (obviously within the limits of the ray theory usage), there is still is an important open question. Standard methods (regular horizontal rays in space) also consider mode coupling in two dimensions, even though such computations appear to be rather complicated \cite{petrov2019application}. However, the next step in the development of our method should be an extension to the description of mode coupling effects (and just their simultaneous propagation) in our three-dimensional space-time. 
\section*{Acknowledgments}
\paragraph{} A. Kaplun is supported by the Center for Integration in Science, Israel Ministry of Aliyah and Integration and post-doctoral scholarship at the Leon H. Charney School of Marine Sciences, University of Haifa.  A. Kaplun is also grateful to V. Kuidin and A.P. Kiselev  for discussions and commentaries. 
\paragraph{}B. Katsnelson is supported by ISF-946/20.

\bibliographystyle{plain}
\bibliography{main}

\end{document}